\documentclass[twocolumn,floats,showpacs,prb]{revtex4}
\usepackage[above]{placeins}
\usepackage{amsmath}
\usepackage{graphicx}

\begin{document}

\title{A unified theory of superconductivity}
\author{Xiuqing Huang$^{1,2}$}
\email{xqhuang@nju.edu.cn}
\affiliation{$^1$Department of Physics and National Laboratory of Solid State
Microstructure, Nanjing University, Nanjing 210093, China \\
$^{2}$ Department of Telecommunications Engineering ICE, PLAUST, Nanjing
210016, China}
\date{\today}

\begin{abstract}
In this paper, we study the reliability of BCS theory as a scientific
explanation of the mystery of superconductivity. It is shown clearly that
the phonon-mediated BCS theory is fundamentally incorrect. Two kinds of
glues, pairing (pseudogap) glue and superconducting glue, are suggested
based on a real space Coulomb confinement effect. The scenarios provide a
unified explanation of the pairing symmetry, pseudogap and superconducting
states, charge stripe order, spin density wave (SDW), checkerboard-type
charge-ordered phase, magic doping fractions and vortex structures in
conventional and unconventional (the high-$T_{c}$ cuprates and MgB$_{2}$)
superconductors. The theory agrees with the existence of a pseudogap in
high-temperature superconductors, while no pseudogap feature could be
observed in MgB$_{2}$ and most of the conventional superconductors. Our
results indicate that the superconducting phase can coexist with a inclined
hexagonal vortex lattice in pure MgB$_{2}$ single crystal with a charge
carrier density $\rho _{s}=1.49\times 10^{22}/cm^{3}.$ Finally, the physical
reasons why the good conductors (for example, Ag, Au, and Cu) and the
overdoped high-$T_{c}$ superconductors are non-superconducting are also
explored.
\end{abstract}

\maketitle

\section{Introduction}

Since the first discovery of superconductivity in mercury in 1911 by H.
Kamerlingh Onnes, \cite{onnes} scientists around the world have been trying
hard to find (or synthesize) the superconducting materials. Through nearly a
century of efforts, it is now clear that superconductivity is an extremely
common natural phenomenon occurring in a wide variety of materials, for
example, pure metals, metallic alloys, heavily-doped semiconductors, a
family of cuprate-perovskite ceramic materials, \cite{bednorz,mkwu} MgB$_{2}$
\cite{nagamatsu} and the newly synthesized iron-based systems. \cite%
{kamihara,xhchen} Soon after the discovery of the superconductivity, the
search for a theoretical understanding of this mysterious phenomenon has
always been one of the hottest topics in condensed matter physics. There are
now thousands of theories on how superconductivity would work but none of
these are definite (including the famous BCS theory \cite{bcs}). The new
experimental evidence in favor of the localized Cooper pairs has just been
reported \cite{stewart}, the discovery shakes the very foundation of the BCS
theory. The new family of superconductors \cite{kamihara} also strongly
challenge the BCS theory based on the electron-phonon coupling mechanism.
\cite{haule,boeri}\ In other words, the mechanism of superconductivity (both
conventional and non-conventional superconductors) remains unsettled. This
raises two questions: (i) What is the main reason of superconductivity in
various superconductors? (ii) Should the mechanisms responsible for
different superconductors be different? In my opinion, any electronic
pairing and superconducting phenomena should share exactly the same physical
reason.

In the earlier works, \cite{huang1} we propose a real space mechanism of
high-$T_{c}$ superconductivity which can naturally explain the complicated
problems, such as pairing mechanism, pairing symmetry, charge stripes,
optimal doping, magic doping fractions, vortex structure, phase diagram,
Hall effect, etc. I am confident that the research may shed light on the
fundamental of superconductivity. In the present paper, we try to extend the
application of the theory in conventional superconductors and MgB$_{2}$ \cite%
{nagamatsu}.

\begin{figure}[tp]
\begin{center}
\resizebox{0.80\columnwidth}{!}{
\includegraphics{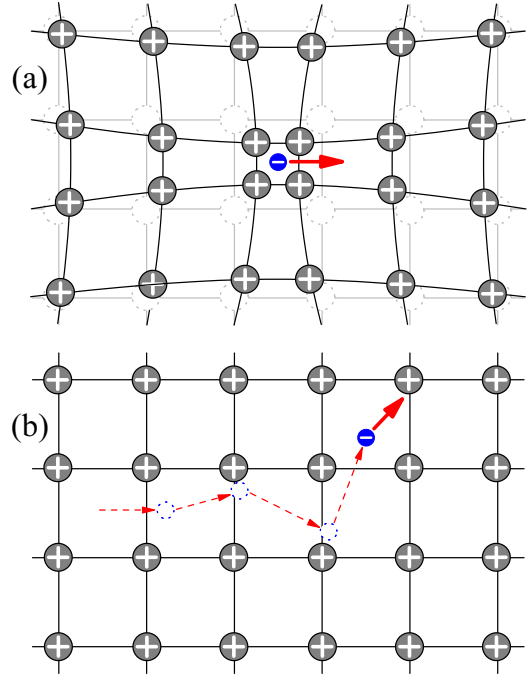}}
\end{center}
\caption{(a) In BCS theory, a single tiny electron can lead to a serious
deformation of lattice structure. (b) We consider that the BCS scenario of
(a) is physically unreasonable, the electron's trajectory will be changed
constantly due to the Coulomb attraction between the electron and ions. }
\label{cooper0}
\end{figure}

\begin{figure*}[tp]
\begin{center}
\resizebox{1.8\columnwidth}{!}{
\includegraphics{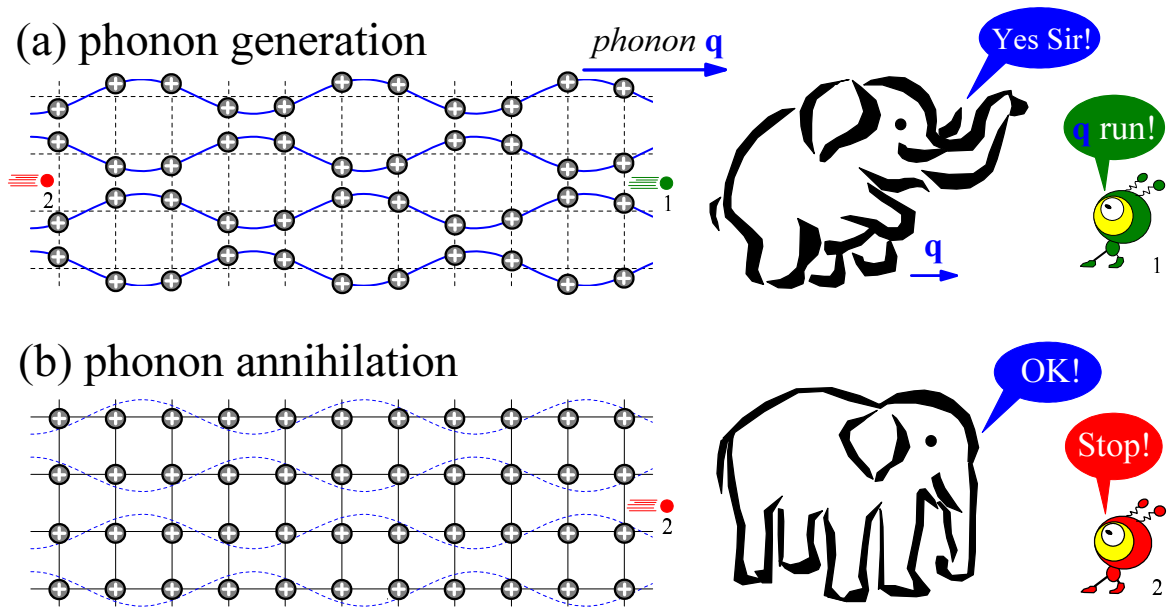}}
\end{center}
\caption{The schematic interpretation of the electron-phonon interaction, or
phonon generation and phonon annihilation inside one Cooper pair based on
the BCS theory. (a) A phonon is generated by \textbf{electron 1}, and (b)
the phonon is annihilated by \textbf{electron 2}. }
\label{phonon}
\end{figure*}

\section{The BCS theory: science or mythology?}

Recently, Anderson pointed out that the need for a bosonic glue (phonon) in
cuprate superconductors is folklore rather than the result of scientific
logic \cite{anderson0}. In this Section, we would like to tell my physics
colleague that the BCS theory is merely a mythology story which even doesn't
work for the conventional superconductors.

\subsection{Who attracts who?}

Fig. \ref{cooper0}(a) shows an electron traveling inside a periodic lattice,
as suggested by BCS, this electron will attract nearby positive charges in
the material. Is this hypothesis physically valid?\

The electron is a fundamental particle that carries a negative electric
charge, and its mass is approximately 1/1836 of that of the proton.
Normally, an atom (or ion) has a mass that is more than 10000 the electron's
mass. All the forces involved in interactions between the electron and ions
can be traced to the electromagnetic interactions. Here, we would like to
raise one question: After exerted by the same amount of force, why the
massive ions get a big-displacement, while the state of the electron remains
almost unchanged. Obviously, the BCS recommended picture of Fig. \ref%
{cooper0}(a) violates the most basic physical principles. A reasonable
physical picture of the electron-ions interactions is presented in Fig. \ref%
{cooper0}(b), where the electron's state is changed constantly and there are
no perceptible changes in the ions's state.

\subsection{\ Can \textquotedblleft Ant\textquotedblright\ command
\textquotedblleft Elephant\textquotedblright ?}

According to BCS theory, a tiny \textbf{electron 1} with a momentum $\mathbf{%
k}$ can cause a collective vibration of the entire lattice (the generation
of a quantized phonon $\mathbf{q}$), as shown in Fig. \ref{phonon}(a). The
scenario implies that, if the entire universe is a single-crystal
superconductor, a single electron can excite a vibration of whole universe.
This sounds like a gigantic \textquotedblleft Butterfly
Effect\textquotedblright . What's even more confusing is when \textbf{%
electron 2} with same $\mathbf{k}$ and opposite spin appears, the phonon
will annihilate instantly, as shown in Fig. \ref{phonon}(b). Or in BCS
language, the phonon is absorbed integrality by \textbf{electron 2}. In
order to continue the story, BCS further assume that the moment the \textbf{%
electron 2} absorbs the phonon, an exact the same phonon has been generated
by the \textbf{electron 1} again, then the new phonon will be absorbed by
\textbf{electron2}, ... Science? or science fiction?

I really don't know why such a absurd theory can become to be the
cornerstone of modern physics. Supposing we now have an \textquotedblleft
ant\textquotedblright\ (electron) and an \textquotedblleft
elephant\textquotedblright\ (atomic lattice) (see Fig. \ref{phonon}), is it
true that the \textquotedblleft ant\textquotedblright\ is the
\textquotedblleft commander\textquotedblright ? Beside, what mechanism can
ensure the leading \textbf{electron 1} never collide with atoms or other
electrons?

\subsection{Do electrons have eyes and ears?}

\begin{figure}[tbp]
\begin{center}
\resizebox{0.95\columnwidth}{!}{
\includegraphics{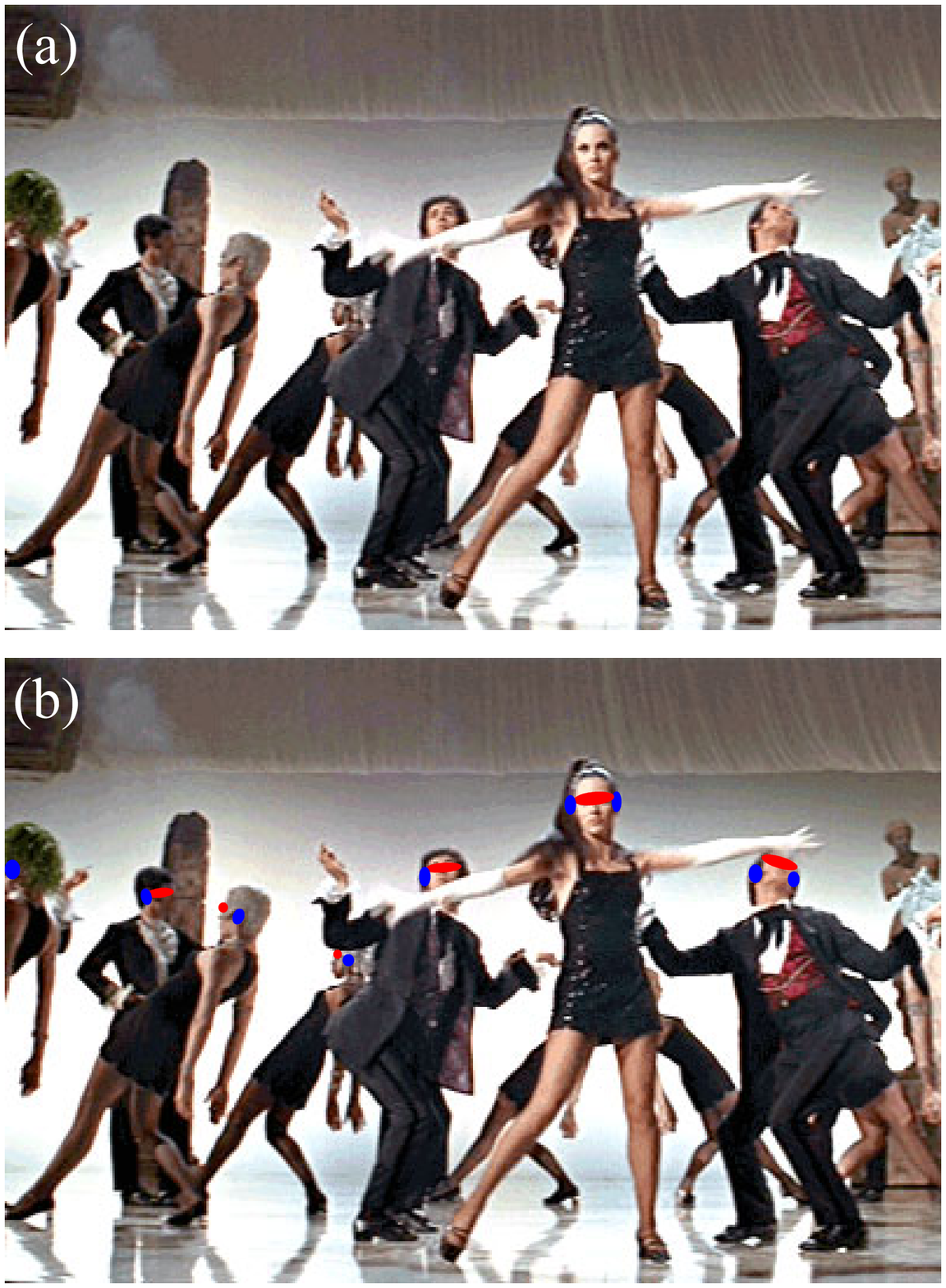}}
\end{center}
\caption{(a) The \textquotedblleft Frug\textquotedblright\ (a popular dance)
which has been applied as an analogy to Cooper pair by Schrieffer. (b)
Obviously, if the dance partners cannot see each other and cannot listen to
the music during the dancing time, one may find the Schrieffer's story quite
ridiculous.}
\label{dance}
\end{figure}

In BCS theory (dynamic screening), the paired electrons are not physically
close together or never in the same place at the same time. It is not clear
how can these extended pairs be crammed together to create a superconducting
medium without getting disrupted. Schrieffer had try to explain how the
loose Cooper pairs can finally lead to the superconductivity. He compared
the concept to the Frug (a popular dance) \cite{frug}, where dance partners
(every male has an up spin and a female has a down spin) could be far apart
and never touch each other (may be a couple of hundred feet apart) on the
dance floor, yet remain a pair, as shown in Fig. \ref{dance} (a).

Although this analogy may sound interesting for audiences lacking of basic
physics knowledge, from the viewpoint of physics, this comparison is
meaningless. Because these two systems are completely different. The main
differences are as follows:

(1) The dance partners have healthy eyes and ears, do electrons have eyes
and ears?

(2) All dancers look different each other, but all electrons are identity.

(3) The dancers are well trained and the choreography well rehearsed, who
have told the electrons how to dance \textquotedblleft
Frug\textquotedblright , God?

(4) In a superconductor, there are many \textquotedblleft
huge\textquotedblright\ (comparing to electron) atomic oscillators. But\
there are not any analogous \textquotedblleft oscillator\textquotedblright\
on the dance floor.

(5) The electrons of the Cooper pair should be momentum opposite, if the
dance partners are also \textquotedblleft momentum\textquotedblright\
opposite, can the dance go on forever?

I'm sure without eyes and ears [see Fig. \ref{dance} (b)], electrons cannot
dance \textquotedblleft Frug\textquotedblright !

\subsection{Equidirectional momentum or opposite momentum?}

\subsubsection{Why two electrons so different?}

\begin{figure}[bp]
\begin{center}
\resizebox{1\columnwidth}{!}{
\includegraphics{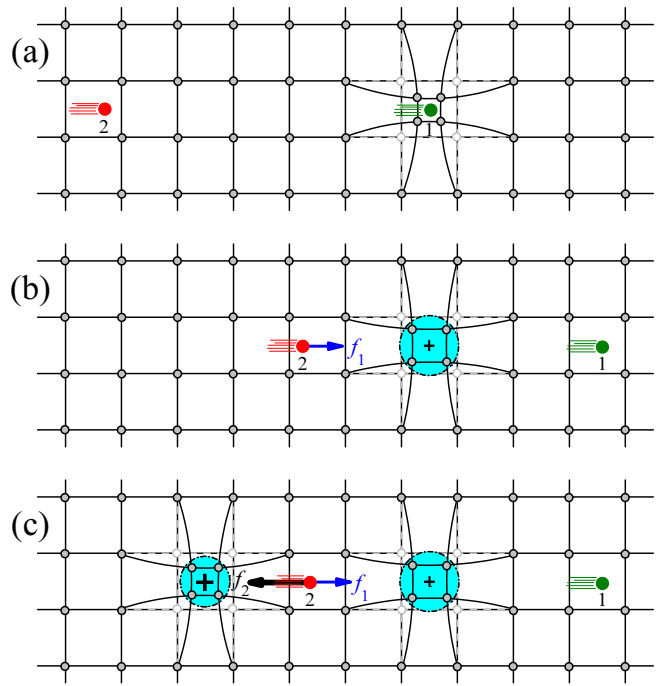}}
\end{center}
\caption{(a)-(b) A traditional model of Cooper pair attraction in real space
where two electrons move in the same direction, (a) a passing \textbf{%
electron 1} attracts the lattice (the positive ions), causing a slight
increase of positive charge center due to Coulomb attraction, (b) and the
trailing \textbf{electron 2} is attracted by it. Here, we argue that the
basic physical pictures described in (a) and (b) are physically untrue. As
seen in the two subfigures, the ripple induced by \textbf{electron 2} has
been completely ignored in this analysis. (c) The actual situation where
both electrons can distort the positively charged ions, independently. In
this case, the phonon induced attraction between the two Cooper pairing
electrons becomes invalid, because the two forces acting on \textbf{electron
2} normally satisfy $f_{2}\gg f_{1}$. }
\label{cooper1}
\end{figure}

\begin{figure}[tbp]
\begin{center}
\resizebox{1\columnwidth}{!}{
\includegraphics{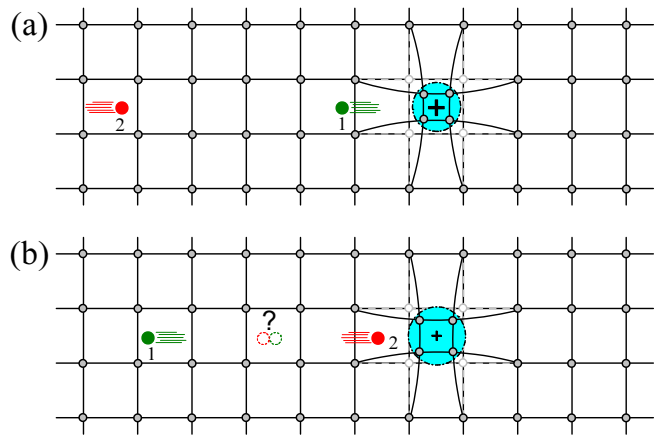}}
\end{center}
\caption{Another visual model of the Cooper pair attraction with two
electrons moving in the opposite direction. (a) The \textbf{electron 1}
distorts the lattice around itself and creates a positive charge density,
(b) the \textbf{electron 2} is attracted to the positive charge center. }
\label{cooper2}
\end{figure}

To represent the \textbf{k}-space's BCS theory in real space, two visual
models of the Cooper pair attraction have been suggested (see Figs. \ref%
{cooper1} and \ref{cooper2}). It is shown here that the real-space
structures of Figs. \ref{cooper1} and \ref{cooper2} cannot follow directly
from the BCS theory and the efforts to explain the condensation of Cooper
pairs are proved to be unreliable or even physically unreasonable. Figs. \ref%
{cooper1} (a) and (b) show the model of Cooper pair attraction in real space
where two electrons move in the same direction \cite{cooper1}. In this case,
the leading \textbf{electron 1} attracts the lattice (the positive ions) and
causes a slight increase of positive charge around it, as shown in Fig. \ref%
{cooper1}(a). This increase in positive charge will, in turn, attract the
trailing \textbf{electron 2}, as shown in Fig. \ref{cooper1}(b). From BCS
theory, we know that this coupling between two electrons is viewed as an
exchange of phonons (the quanta of lattice vibration energy). However, this
real space picture is totally inconsistent with the \textbf{k}-space's BCS
theory in many aspects. As is well known, the BCS theory asserts that the
two paired electrons must have opposite spin and opposite momentum. But,
Figs. \ref{cooper1} (a) and (b) show clearly that the real space
representation of the bound Cooper pair electrons are in the same momentum.
Furthermore, this approach fails to explain why the two paired electrons
should be spin antiparallel.

In fact, the major flaw of the BCS theory is that the lattice distortion
caused by \textbf{electron 2} has been completely ignored in this analysis.
In our opinion, a complete picture of the real space description of BCS
theory must take into account not only the \textbf{electron 1} but also the
\textbf{electron 2}, as illustrated in Fig. \ref{cooper1}(c). From this
figure it is clear that there are two forces acting on \textbf{electron 2}:
the attractive force $f_{1}$ produced by the positive charge center of
\textbf{electron 1} and the drag force $f_{2}$ exerted by the positive
charge center of \textbf{electron 2} itself. BCS theory suggests that
electron pairs can couple over a range of hundreds of nanometers, three
orders of magnitude larger than the lattice spacing, therefore, the drag
force $f_{2}$ is generally much larger than the attractive force $f_{1}$.
This further implies phonon-mediated BCS theory is not valid in physics.

\subsubsection{How to avoid electron-electron repulsion and how to maintain
a instantaneous \textquotedblleft attraction\textquotedblright\ forever?}

Figure \ref{cooper2} shows another visual model of the Cooper pair
attraction with two electrons moving in the opposite direction \cite{cooper2}%
. Compared with Figures \ref{cooper1}, although now the two electrons have
opposite momentum as suggested by BCS theory, apart from the spin and phonon
issues discussed above, there are a number of fatal problems with this
explanation. First, even if the positive charge center of Fig. \ref{cooper2}
(a) can attract another electron passing in the opposite direction [see Fig. %
\ref{cooper2} (b)], apparently, the attraction is instantaneous. Second,
when two electrons approach each other, a strong electron-electron repulsion
is unavoidable [Fig. \ref{cooper2} (b)]. All these factors indicate the
Cooper pair should split up rather than stay together when the pair is
formed by two electrons with opposite momentum.

In a word, two real space electron-phonon mechanisms are examined and they
cannot give a satisfactory explanation of the BCS theory. If BCS still
cannot provide a convincing real space picture of the phonon-mediated BCS
theory, thus there is good reason to doubt: Is BCS theory correct?

\section{Real space correlation between two parallel-spin electrons}

\begin{figure}[tp]
\begin{center}
\resizebox{1\columnwidth}{!}{
\includegraphics{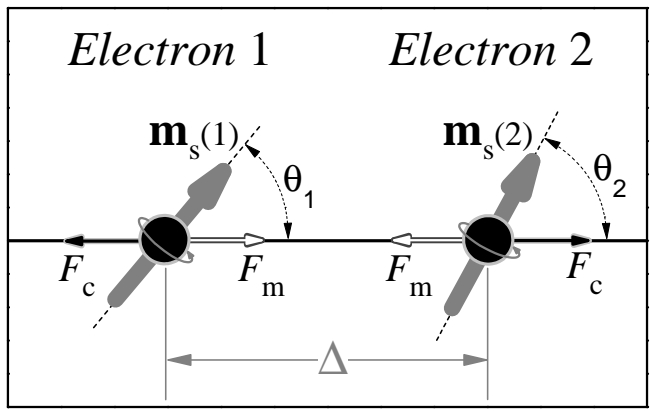}}
\end{center}
\caption{The electromagnetic interaction between two electrons. Normally,
there is a pair of long-range Coulomb repulsion $F_{c}$. In addition, two
spinning electrons may create a pair of short-range attractive forces $F_{m}$
due to the dipolar-dipolar interaction.}
\label{dipolar}
\end{figure}

Since the observation of real-space ordering of charge in cuprate
superconductors, \cite{kivelson,tranquada,norman,ichikawa} it is widely
accepted that short-range electron-electron correlations can bind electrons
into real space pairs and dominate the superconductivity properties of the
materials. Normally, as shown in Fig. \ref{dipolar}, for the two static
electrons, there is a long-range repulsive electron-electron Coulomb
interaction
\begin{equation}
F_{c}=\frac{e^{2}}{4\pi \varepsilon _{0}\Delta ^{2}},  \label{fc}
\end{equation}%
where $e$ is the electron charge and $\Delta $ is the distance between two
electrons.

It is known that study of superconducting correlations in conventional
superconductors is always performed in momentum-space (dynamic screening), \
where the paired electrons are seldom or never in the same place at the same
time. \cite{anderson0} In the case of dynamic screening, only the long-range
Coulomb interaction $e^{2}/\Delta $ is considered while the short-range
electron--electron magnetic interactions is completely ignored. We argue
here that, in the case of real-space screening, the magnetic forces among
the electrons (see also Fig. \ref{dipolar}) should be taken into account.
Approximately, the magnetic dipolar interaction forces $F_{m}$ exerted on
the electrons are given by
\begin{equation}
F_{m}\approx \frac{3\mu _{0}\mu _{B}^{2}}{2\pi \Delta ^{4}}\cos \theta
_{1}\cos \theta _{2},  \label{fm}
\end{equation}%
where $\mu _{0}$ is the permeability of free space and $\mu _{B}$ is the
Bohr magneton.

The forces $F_{m}$ of Eq. (\ref{fm}) can be attractive and repulsive
depending on the orientation ($\theta _{1}$and $\theta _{2}$) of electron
magnetic moment $\mathbf{m}_{s}(j)$, ($j=1,2$). When $\theta _{1}=\theta
_{2}=0$ (or $\pi )$, the magnetic poles of the paired electrons are lined up
in parallel (spin-parallel pair correlation), contrary to the spin
antiparallel BCS theory. Consequently, the attractive magnetic force reaches
its maximum value $F_{m}^{\max }=3\mu _{0}\mu _{B}^{2}/2\pi \Delta ^{4}$,
and the electron pair corresponds to the most stable and energy minimum
state. When $\left\vert \theta _{1}-\theta _{2}\right\vert =$ $\pi ,$ the
two electrons are spin antiparallel obeying the BCS theory, but the
corresponding pair is in the most unstable and maximum energy state. Such a
construal has significant implication that the BCS theory is physically
unreasonable.

\begin{figure}[bp]
\begin{center}
\resizebox{0.95\columnwidth}{!}{
\includegraphics{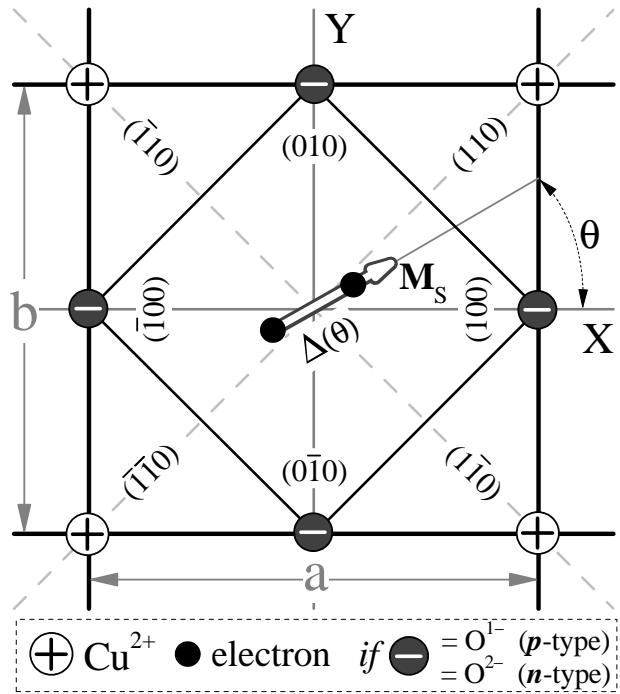}}
\end{center}
\caption{Two spin parallel electrons with a joint magnetic moment $\mathbf{M}%
_{s}$ is confined inside one unit cell of CuO plane. Only nearest-neighbor
(four oxygen ions) and next-nearest-neighbor (four copper ions) interactions
are considered. The nearest-neighbor negative charge of the oxygen irons
play a key role for the pseudogap phenomenon in cuprates.}
\label{confine}
\end{figure}

\section{ Pairing glue and pseudogap}

At high temperatures, the vibrational motion of the material's lattice
becomes so stiff that it tends to break up the electron pairs instead of
holding them together. \cite{anderson0} So what could possibly provide the
glue that keeps the carriers bound in Cooper pairs? Although many candidates
for this glue (including spin fluctuations, phonons, polarons, charge
stripes and spin stripes) have been proposed, what the pairing glue in high-$%
T_{c}$ cuprates is still an open question. In this Section, we would like to
discuss the issue from the point of view of real-space confinement effect.
To describe this, two spin parallel electrons of Fig. \ref{dipolar} with a
joint paired-electron magnetic moment $\mathbf{M}_{s}=\mathbf{m}_{s}(1)+%
\mathbf{m}_{s}(2)=2\mathbf{m}_{s}$ are embedded into a CuO plane of the
cuprate superconductor, as shown in Fig. \ref{confine}.

Looking at the figure, just a simplification, only nearest-neighbor and
next-nearest-neighbor interactions are considered.\ Inside the unit cell,
the possible paired-electrons with the magnetic moment $\mathbf{M}_{s}$
along the $\theta $ direction, and the corresponding distance between the
electrons is reexpressed as $\Delta (\theta )$. From the figure, one can
easily conclude that the pair with the $\mathbf{M}_{s}$ oriented in (100),
(010), ($\overline{1}$00) and (0$\overline{1}$0) directions is generally
considered to be much more stable (minimum energy) due to the suppression of
the four oxygen ions (O$^{1-}$ for hole-doped, O$^{2-}$ for electron-doped),
as opposed to the cases, in (110), ($\overline{1}$10), ($\overline{1}%
\overline{1}$0) and (1$\overline{1}$0) directions where the bound pair tends
to be separated by Coulomb forces of the Cu$^{2+}$. As a consequence, the
distance $\Delta (\theta )$ between the two electrons of the pair has a
minimum (maximum binding energy) at $\theta =0$, $\pi /2$, $\pi $ and 3$\pi
/2$, while at $\theta =\pi /4$, $3\pi /4$, $5\pi /4$ and $7\pi /4$, $\Delta
(\theta )$ will reach its maximum value (minimum binding energy). Obviously,
the unified model (see Fig. \ref{confine}) for both hole- and electron-doped
cuprates has essentially the same pairing mechanisms (pairing glue). In the
previous paper, \cite{huang1} a more detailed study was done based on the
Coulomb's equation and the results suggested the dominant $d$-wave symmetry
in hole-doped cuprates and a possible mixed $(s+d)$-wave symmetry in
electron-doped systems. The results revealed that the localized
electromagnetic interactions are indeed the source (glue) of localized
cooper pairs characterized by the pseudogap.

\begin{figure*}[tbp]
\begin{center}
\resizebox{1.6\columnwidth}{!}{
\includegraphics{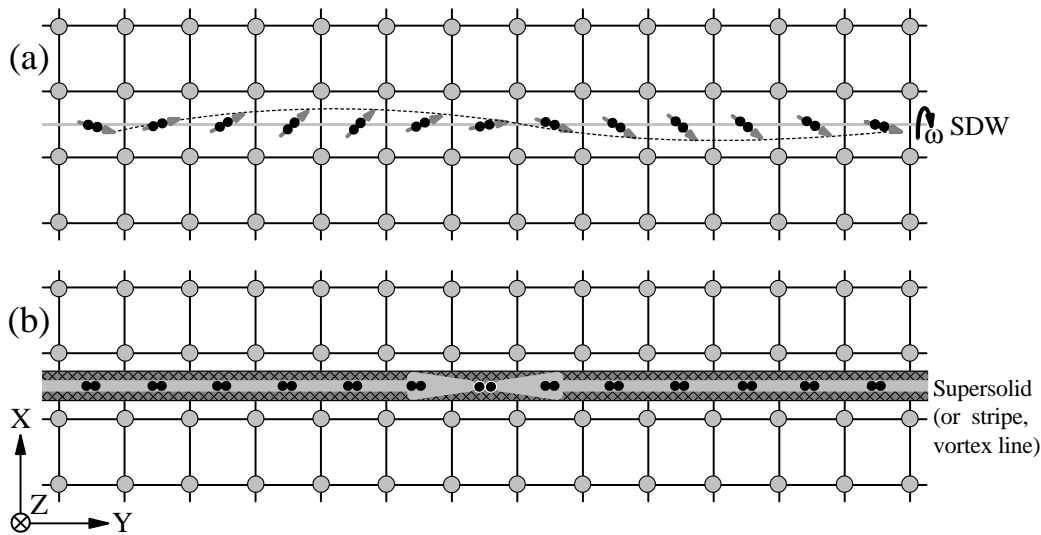}}
\end{center}
\caption{(a) Due to the magnetic phase-coherence among the electron pairs, a
helical dynamical spin-density-wave (SDW) is inspired in the metallic charge
stripe (vortex line) and the superconductivity and SDW can coexist along
this stripe, (b) when the electron pairs are highly coherent, the charge
stripe can be considered as a `supersolid' where any electron pair inside
always experiences a pair of compression forces (a repulsive force pairing
mechanism). If background ions are positive, pairing and superconducting
will occur at the same time.}
\label{confine1}
\end{figure*}

The nature of the normal-state gap (pseudogap) phase of HTSC is still highly
controversial. ARPES and tunneling measurements show a clear pseudogap which
was seen to persist even at room temperature. \cite{loeser,renner,ding2}
There are many models attempt to describe the mysterious pseudogap state.
Strictly speaking, none of the proposed models is completely satisfactory.
As discussion above, here we present a new approach based on the simple and
natural picture of the real-space confinement effect, and the pseudogap is
associated with the local structure of unit cell in CuO$_{2}$ plane. Thus it
should not be surprising about the pseudogap behavior which indicate the
formation of pairs (localized cooper pairs) below $T^{\ast }>T_{c}.$

In our viewpoint, pseudogap phenomenon is merely a real space confinement
effect in the superconductors if electrons were confined inside one unit
cell. Besids, to have a stable pseudogap phase, pair-pair interactions
should be suppressed. Hence, decreasing the charge carrier density is a
useful way to open and maintain a pseudogap in the superconductors. In other
words, the pseudogap is associated with the local structure and the charge
carrier density in the superconductors. Therefore, the localized Cooper
pairs (pseudogap) are likely to survive in insulating or nonmetallic
materials \cite{stewart}.

\section{ A collective confinement and superconducting glue}

Physically, pairing in cuprates is an individual behavior characterized by
pseudogap, while superconductivity is a collective behavior of many coherent
electron pairs. Nowadays, more and more beautiful experimental results
suggest that stripes are common in cuprates and may be important in the
mechanism for HTSC. In the paper, \cite{huang1} based on the GL theory
formalism, we argued that the dimerized charge stripes can contribute to the
mechanism of superconductivity in cuprate superconductors and the dynamical
spin density wave (SDW) coherent phases can be established along the
stripes. As can be seen, the high-$T_{c}$ superconducting order is also
inherently related to the a real space collective confinement .
Consequently, the superconductivity has an origin different from pseudogap
in high-$T_{c}$ superconductors. Here a similar real space collective
confinement picture is introduced into the conventional superconductors, as
shown in Fig. \ref{confine1}, very different from the localized pairing
mechanism (see Fig. \ref{confine}) of the high-$T_{c}$ superconductors. In
this case, a real space helical dynamical spin-density-wave [ Fig. \ref%
{confine1} (a)] and superconductivity coexist to form a dimerized charge
supersolid (a charge-Peierls dimerized transition), as shown in Fig. \ref%
{confine1} (b). Indeed, both the pairing and superconducting (phase
coherence) occur simultaneously at $T_{c}$, as generally accepted
experimental facts. In the real space collective confinement picture,\ the
so-called spin density wave (SDW), superconducting charge stripe and the
vortex line are exactly the same thing. Anyway, the spin correlation of Fig. %
\ref{confine1} is a general phenomenon in superconductors, and it must be
the fundamental to the mechanism (superconducting \textquotedblleft
glue\textquotedblright ) of superconductivity in conventional and
unconventional superconductors.

\section{ La$_{2-x}$Sr$_{x}$CuO$_{4}$}

\begin{figure}[tbp]
\begin{center}
\resizebox{0.9\columnwidth}{!}{
\includegraphics{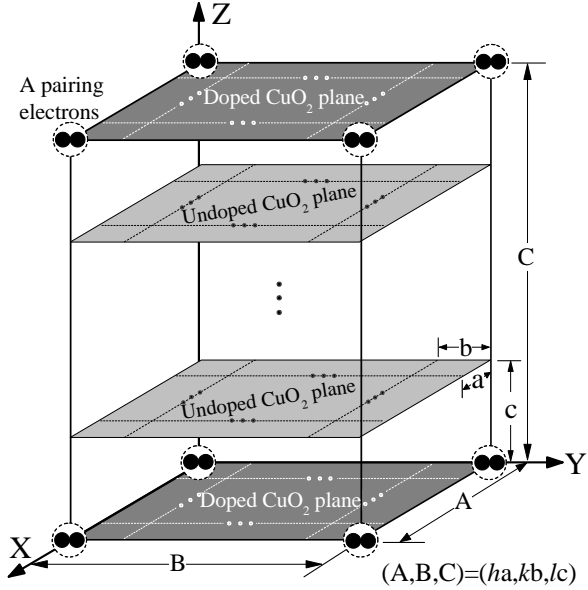}}
\end{center}
\caption{Simplified schematic unitcell of the electron-pairs (dimerized)
Wigner crystal in the high-$T_{c}$ cuprates.}
\label{wigner}
\end{figure}

In nature, periodic structures are often considered as the result of
competition between different interactions. The formation of stripe patterns
is generally attributed to the competition between short-range attractive
forces and long-range repulsive forces. \cite{seul} In the paper,\cite%
{huang1} we argued that, in the proper doped LSCO superconductor, the
electron pairs can self-organize into a `superlattice' (Wigner crystal of
electron pairs) with the primitive cell $(A,B,C)=(ha,kb,lc)$, as shown in
Fig. \ref{wigner}. Consequently, the \textquotedblleft
material\textquotedblright\ composed of electron-pair \textquotedblleft
atoms\textquotedblright\ will undergo a structure transition from random to
order phase (LTO, LTT). Thus, the doping level $x$ is given by

\begin{equation}
x=p(h,k,l)=2\times \frac{1}{h}\times \frac{1}{k}\times \frac{1}{l},
\label{fractions}
\end{equation}%
and the corresponding charge carrier density is%
\begin{equation}
\rho _{s}=\frac{2}{ABC}=\frac{2}{hkl}\frac{1}{abc}=\frac{x}{abc},
\label{density}
\end{equation}%
where $h$, $k$, and $l$ are integral numbers. Note that, from the viewpoint
of energy, it is also possible that the `superlattice' exhibits two simple
hexagonal structures (see below).

\subsection{LTT1($h,k,l$) non-superconducting phase}

\begin{figure}[tbp]
\begin{center}
\resizebox{1\columnwidth}{!}{
\includegraphics{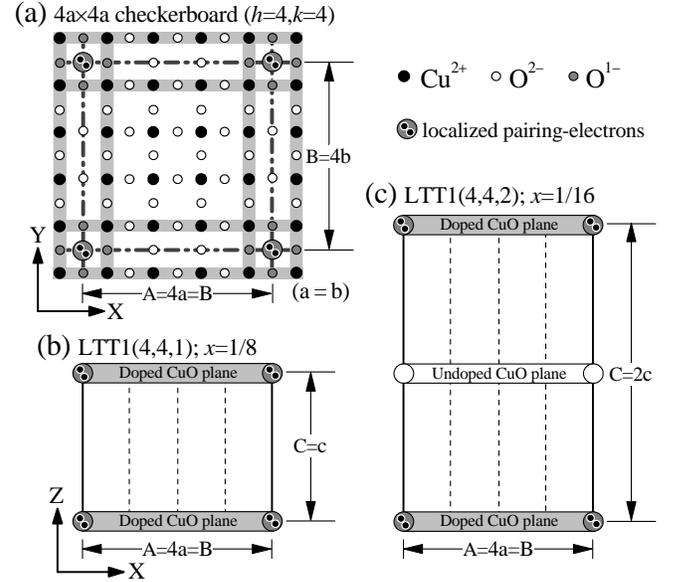}}
\end{center}
\caption{The nondispersive superlattices of the electron pairs in the LSCO.
(a) The $4a\times 4a$ checkerboard in the doped CuO$_{2}$ planes, when
doping levels at $x=1/8$ or $x=1/16$. (b) LTT1(4,4,1) phase where all CuO$%
_{2}$ planes are doped. (c) LTT1($4,4,2$) phase, only half of the CuO$_{2}$
planes are doped. }
\label{ltt1_8}
\end{figure}

We found that there are only five abnormal phases (the so-called
\textquotedblleft magic doping phases\textquotedblright ) in LSCO, which are
related to the anomalous suppression of superconductivity. They are LTT1($%
6,6,1$) of $x=1/18$ , LTT1($4,4,1$) ($x=1/8$), LTT1($4,4,2$) ($x=1/16$),
LTT1($3,3,2$) ($x=1/9$) and LTT1($2,2,2$) ($x=1/4$) where the nondispersive
superlattices of $6a\times 6a$, $4a\times 4a$, $4a\times 4a$, $3a\times 3a$
and $2a\times 2a$ in CuO$_{2}$ planes can be expected, respectively. At $%
x=1/8,$ LTT1($4,4,1$) can also coexists with the LTT original lattice ($a=b$%
) of the LSCO [see Fig. \ref{ltt1_8}(a) and (b)]. This may explain the
famous \textquotedblleft 1/8 anomaly\textquotedblright\ in various high-$%
T_{c}$ superconductors. \cite%
{tranquada,valla,moodenbaugh,crawford,homes,fujita} Note that although the
nondispersive $4a\times 4a$ superstructure [see Fig. \ref{ltt1_8}(a)] seems
to be exactly the same in both samples ($x=1/8$ and 1/16). \cite%
{hanaguri,kim} We show, for the first time, that two samples are in fact
very different: in the sample of $x=1/8$ indicated by LTT1($4,4,1$) in this
paper, where all CuO$_{2}$ planes are doped [Fig. \ref{ltt1_8}(b)]; while at
$x=1/16$ of LTT1($4,4,2$), only half of the CuO$_{2}$ planes (every two
planes) are doped [Fig. \ref{ltt1_8}(c)].

Encouragingly, apart from the $x=1/8$, some unusual results have already
been observed at $x=1/16$, $x=1/9$ and $x=1/4$ of the doped LSCO crystals.
For instance, by high resolution ARPES experiments on $x\sim 1/16$ sample,
an anomalous change at $\sim 70$ mev in the nodal scattering rate was
reported, \cite{zhou1} and the observations of intrinsic anomalous
superconducting properties at magic doping levels of $x=1/16$ and $x=1/9$
had been found by $dc$ magnetic measurements. \cite{zhou2} The experimental
verification of the strong-correlation fluctuations in a non-superconductive
$x=1/4$ sample has been noted. \cite{goodenough} Most recently, Wakimoto
\textit{et al. }\cite{wakimoto1} reported the structural and
neutron-scattering experiment study for over-doped LSCO with $x=1/4.$ They
confirmed that the crystal structure of the composition has tetragonal
symmetry (LTT1 phase) with lattice constant of $a=b=3.73$ $\mathring{A}$ at
10 K and the IC peaks appear around the antiferromagnetic wave vector $%
(1/2,1/2$). These facts would add considerable support that our theory has
the great merit of explaining high-$T_{c}$ superconductivity.

\subsection{LTT2($h,k,l$), LTT3($h,k,l$), SH1($h,k,l$) and SH2($h,k,l$)
superconducting phases (vortex lattices)}

\begin{figure}[tbp]
\begin{center}
\resizebox{1\columnwidth}{!}{
\includegraphics{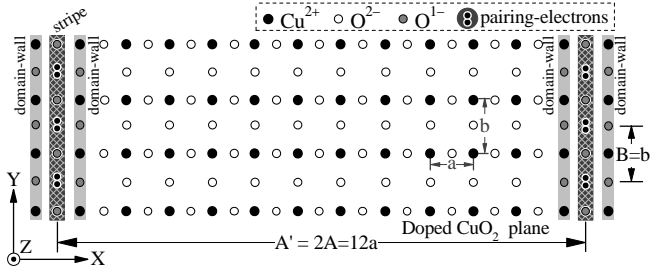}}
\end{center}
\caption{Periodic dimerized charge stripes (vortex lines) in doped CuO$_{2}$
of LSCO.}
\label{lsco1}
\end{figure}

\begin{figure}[tbp]
\begin{center}
\resizebox{1\columnwidth}{!}{
\includegraphics{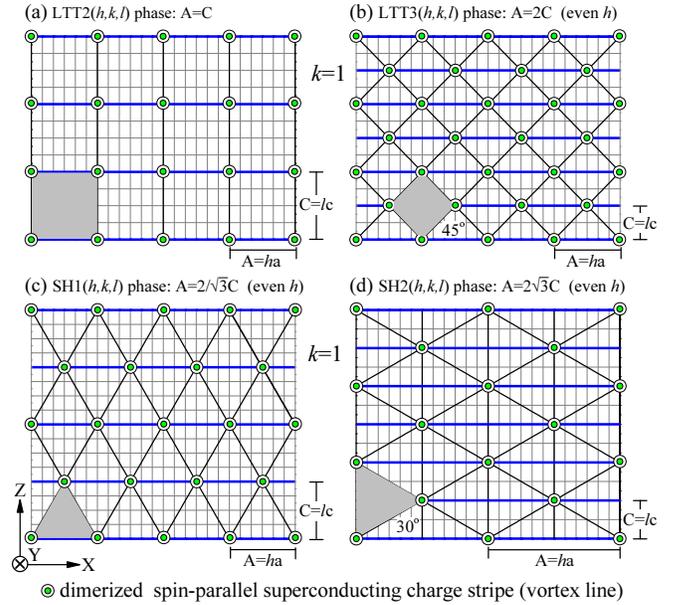}}
\end{center}
\caption{Four quasi-two-dimensional vortex lattices with a uniform
distribution of vortex lines. (a) LTT2($h,k,l$) phase, the charge stripes
have a tetragonal symmetry in XZ plane, (b) LTT3($h,k,l$), the vortex
lattice has a tetragonal symmetry in XZ plane with a orientation 45$^{0}$.
(c) and (d) The simple hexagonal (SH) phases [SH1($h,k,l$) and SH2($h,k,l$%
)]. }
\label{4phase}
\end{figure}

According to Fig. \ref{wigner} and Eq. (\ref{fractions}), the metallic
charge stripes (vortex lines) are periodic spatial modulations\ in the doped
CuO$_{2}$ planes (XY plane) of the LSCO, as shown in Fig. \ref{lsco1}. But
what concerns us here still more is the charge-stripe order in the XZ plane
perpendicular to the plane of CuO$_{2}.$We argue that the physically
significant critical value for the stable charge-stripe order is that at
which $T_{c}$ is maximum. In this sense, the LTT2, LTT3 and the simple
hexagonal (SH) phases (vortex lattices) might be the ideal candidates for
the stable charge-stripe order of paired electrons. In the LTT2($h,k,l$)
phase, as shown in Fig. \ref{4phase} (a), the charge stripes have a
tetragonal symmetry in XZ plane in which the superlattice constants satisfy
\begin{equation}
\frac{A}{C}=\frac{ha}{lc}=1.  \label{LTT2}
\end{equation}%
Fig. \ref{4phase} (b) shows the LTT3($h,k,l$), the vortex lattice has a
tetragonal symmetry in XZ plane with a orientation 45$^{\text{0}}$ and the
superlattice constants:
\begin{equation}
\frac{A}{C}=\frac{ha}{lc}=2.  \label{LTT3}
\end{equation}%
While in simple hexagonal (SH) phases, as shown in Figs. \ref{4phase} (c)
and (d), the charge stripes possess identical trigonal crystal structures.
In the SH1($h,k,l$) phase [see Fig. \ref{4phase} (c)], the superlattice
constants have the following relation
\begin{equation}
\frac{A}{C}=\frac{ha}{lc}=\frac{2\sqrt{3}}{3}\approx 1.154700.  \label{SH1}
\end{equation}%
For the SH2($h,k,l$) phase of Fig. \ref{4phase} (d), this relation is given
by
\begin{equation}
\frac{A}{C}=\frac{ha}{lc}=2\sqrt{3}\approx 3.46410.  \label{SH2}
\end{equation}

\begin{figure}[tbp]
\begin{center}
\resizebox{0.95\columnwidth}{!}{
\includegraphics{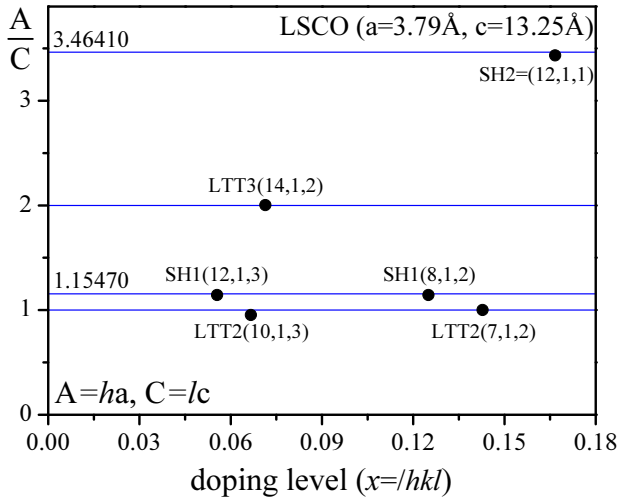}}
\end{center}
\caption{{}Based on the experimental lattice constants, the candidates for
the optimal doping and the superconducting phases in LSCO system are given.
We consider that the maximum high-$T_{c}$ phase (optimal doping) may be
relevant to SH2($12,1,1$) with the optimal doping density $x=1/6\approx
0.1667$.}
\label{lsco2}
\end{figure}

It is commonly accepted that samples of La$_{2-x}$Sr$_{x}$CuO$_{4}$ have the
highest $T_{c}$ at Sr concentration (optimal doping) $x\sim 0.16\ $with the
experimental lattice constants: $a=3.79\mathring{A}$ and $c=13.25\mathring{A}%
.$ In this subsection, basing on the above analysis, we will attempt to
provide a general description of the stable superconducting phase (metallic
stripe) in LSCO and give a possible relationship between the lattice
constants and optimal doping phase. Note that although the LTT2 and SH
phases have rather different spatial structure, Eqs. (\ref{fractions}) and (%
\ref{density}) are still valid for the SH phases due to the appropriate
definition of the superlattice constants ($A$ and $C$) in Figs. \ref{4phase}
(b) and (c). In LSCO, based on the experimental lattice constants ($a=3.79%
\mathring{A}$ and $c=13.25\mathring{A}),$ the relationship between the
doping level $x=/1hkl$ and $A/C$ is shown in Fig. \ref{lsco2}. We argue that
the SH1($12,1,3$) vortex phase of $x=1/18\approx 0.05555$ is most likely the
lowest doped superconducting LSCO sample, in favor of the experimental
result that superconductivity emerges at $x\sim 0.056$ in LSCO
superconductor \cite{wakimoto,keimer,kastner}. The superconducting SH1($%
8,1,2 $) ($x=1/8)$ is completely suppressed by the non-superconducting
phases of LTT1($4,4,1$) ($x=1/8$). We find several candidates for the
optimal doping phase in LSCO system (see Fig. \ref{lsco2}). Among the
residual phases [SH2($12,1,1$) of $x=1/6$ and LTT2($7,1,2$) of $x=1/7$], we
consider that the maximum high-$T_{c}$ phase (optimal doping) may be
relevant to SH2($12,1,1$). Using Eqs. (\ref{fractions}) and (\ref{density}),
one arrives at the analytical values of the optimal doping density $%
x=1/6\approx 0.1667$ and charge carrier density $\rho _{s}\sim 8.76\times
10^{20}/cm^{3},$ in reasonable agreement with the experiments ($x\sim 0.16$
and $\rho _{s}\sim 9\times 10^{20}/cm$). Two other superconducting phases
LTT2($10,1,3$) of $x=1/15$ and LTT3($14,1,2$) of $x=1/14$ are also
analytically determined.

\subsection{Phase diagram}

\begin{figure}[tbp]
\begin{center}
\resizebox{1\columnwidth}{!}{
\includegraphics{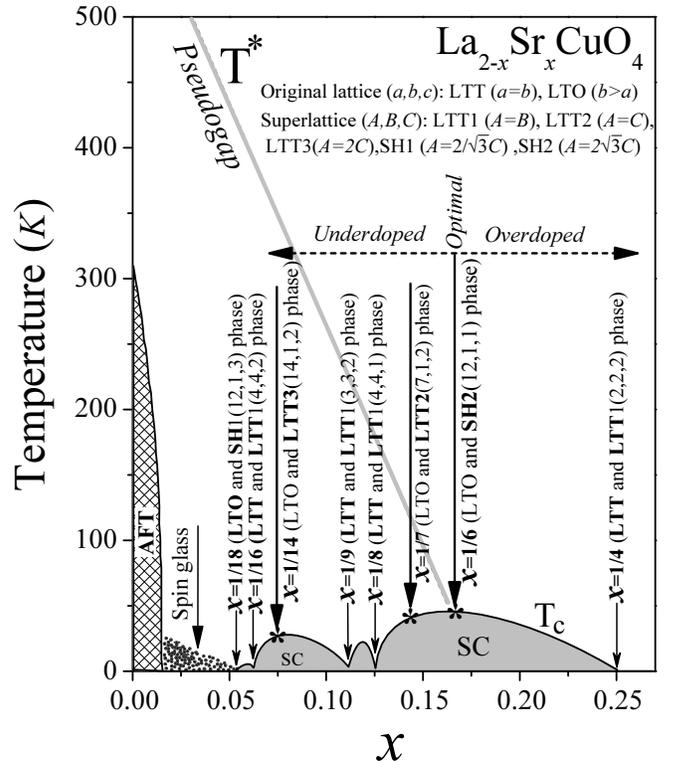}}
\end{center}
\caption{An analytical phase diagram for LSCO. There are four abnormal
phases (at $x=1/4$, 1/8, 1/9 and 1/16), where the LTT1 superlattice phases ($%
A=B$) can coexist with the LTT original lattices ($a=b$) in the LSCO. And
the optimal doping phase ($x=1/6$), where the metallic SH2($12,1,1$)
charge-stripe phase can be expected. Moreover, several new superconducting
phases [for example, SH1($12,1,3$) and LTT2($7,1,2$)] are also predicted by
our theory.}
\label{phase}
\end{figure}

From the discussion of our results, we summarize the doping dependence of $%
T_{c}$ for LSCO in a schematic phase diagram in Fig. \ref{phase}. It is well
known that the antiferromagnetic Mott insulator phase is found near the
origin of La$_{2}$CuO$_{4}$. For doping beyond a few percent, the material
enters the disordered phase (spin glass). At $x=1/18$, the material will
undergo an insulator-to-metal transition, at the same time displaying
superconductivity at low temperature. According to Eq. (\ref{fractions}),
the \textquotedblleft magic effect\textquotedblright\ \cite{moodenbaugh} is
possibly taking place at rational doping levels 1/4, 1/8, 1/9, 1/16 and
1/18, where the LTT1 superlattice phases ($A=B$) can coexist with the LTT
original lattices ($a=b$) in the LSCO. In these specific situations, the
paired electrons are localized, hence the corresponding charge orders appear
to be completely destructive to superconductivity.

We note here that the bosonic theory predicts all magic doping fractions at $%
x=(2m+1)/2^{n},$ where $m$ and $n$ are integers, \cite{komiya} which implies
the possibility of an infinite magic doping fractions in LSCO, while our
theory predicts commensurate effect only at four magic doping fractions 1/4,
1/8, 1/9 and 1/16 (see Fig. \ref{phase}). The reported measurements find a
tendency towards charge ordering at four particular rational doping
fractions of 1/4, \cite{wakimoto1} 1/8, \cite%
{tranquada,moodenbaugh,crawford,homes,fujita} 1/9, \cite{zhou1,zhou2} and
1/16 \cite{kim} and is most consistent with our theoretical prediction. In
view of the intriguing agreement of the experimental data with our model, it
would be desirable to systematically perform direct measurements of the
charge order in the underdoped LSCO materials, where the nondispersive
checkerboard-type ordering with periodicity $3a\times 3a$ can be
experimentally observed at the doping level $x=1/9$.

\begin{table*}[th]
\caption{Lattice constants, charge carrier density, the optimal doping
levels (analytical values $x=2/hkl$ and experimental data $x^{\prime }$) and
the possible optimal superconducting charge-stripe phases (vortex lattices)
in cuprate and MgB$_{2}$.}
\label{table1}%
\begin{tabular}{ccccccccccccc}
\hline\hline
Superconductors & $a($\AA $)$ & $a^{\prime }($\AA $)$ & $b($\AA $)$ & $c($%
\AA $)$ & $h$ & $k$ & $l$ & $A/C$ & Vortex phase & $\rho _{s}(cm^{-3})$ & $x
$ & $x^{\prime }$ \\ \hline
La$_{2-x}$Sr$_{x}$CuO$_{4}$ & 3.79 &  & 3.80 & 13.25 & 12 & 1 & 1 & 3.433 &
SH2(12,1,1) & $8.73\times 10^{20}$ & $1/6\approx 0.1667$ & $0.16$ \\
&  &  &  &  & 7 & 1 & 2 & 1.001 & LTT2(7,1,2) & $7.48\times 10^{20}$ & $1/7$
&  \\
YBCO$_{124}$ & 3.84 &  & 3.87 & 27.24 & 7 & 1 & 1 & 0.988 & LTT2(7,1,1) & $%
7.06\times 10^{20}$ &  &  \\
YBCO$_{247}$ & 3.85 &  & 3.87 & 50.29 & 13 & 1 & 1 & 0.995 & LTT2(13,1,1) & $%
3.04\times 10^{20}$ &  &  \\
BSCCO$_{2212}$ & 3.80 &  & 3.80 & 30.80 & 8 & 1 & 1 & 0.989 & LTT2(8,1,1) & $%
7.33\times 10^{20}$ &  &  \\
BSCCO$_{2223}$ & 3.80 &  & 3.80 & 37.82 & 10 & 1 & 1 & 1.004 & LTT2(10,1,1)
& $3.66\times 10^{20}$ &  &  \\
TBCO$_{2201}$ & 3.90 &  & 3.90 & 23.20 & 6 & 1 & 1 & 1.008 & LTT2(6,1,1) & $%
9.44\times 10^{20}$ &  &  \\
MgB$_{2}$ & 3.086 &  &  & 3.524 & 2 & 1 & 2 & 0.876 & SH2(2,1,2) & $%
1.49\times 10^{22}$ &  &  \\ \hline\hline
\end{tabular}%
\end{table*}

While at $x=1/15$, $1/14$, 1/7 and 1/6, the stable quasi-one-dimensional
metallic charge stripe orders can coexist with superconductivity.
Consequently, the high $T_{c}$ stable superconducting phases (vortex
lattices) are associated with the special doping levels (1/15, 1/14, 1/7 and
1/6) of LSCO, as shown in Fig. \ref{lsco2} and Fig. \ref{phase}.

\section{Vortex lattices in cuprate, MgB$_{2}$ and pure metallic
superconductors}

The existence of LTT2 and SH charge-stripe phases in superconductors is most
likely a universal feature as shown clearly in Table \ref{table1}. We
believe that there is an intrinsic relationship between the vortex structure
and the LTT2 and SH charge-stripe phases of superconductors, in other words,
they are exactly the same thing. This (see Fig. \ref{4phase}) may explain
why in some cases the Abrikosov flux lattices \cite{abrikosov} are
experimentally observed in conventional type II superconductors, \cite%
{essmann,trauble} high-$T_{c}$ superconductors \cite{gammel,bolle} and MgB$%
_{2}.$ \cite{eskildsen} From these data, it becomes evident that both
hexagonal and square vortex lattices can be observed in many conventional
and non-conventional superconductors. \cite{brown} In BSCCO$_{2212}$, the
results suggest a possible tetragonal LTT2($8,1,1$) phase which may explain
the observation of short-range vortex phase having square symmetry. \cite%
{matsuba}
\begin{figure}[bp]
\begin{center}
\resizebox{1\columnwidth}{!}{
\includegraphics{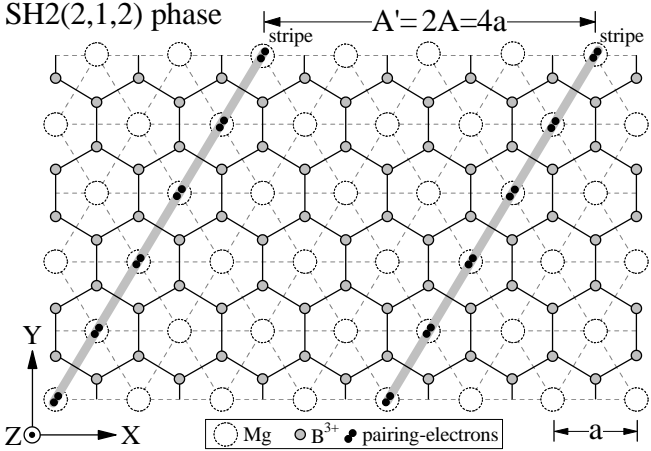}}
\end{center}
\caption{The periodic stripes (vortex lines) in superconducting B-plane of
MgB$_{2}$ superconductor. }
\label{mgb1}
\end{figure}

In MgB$_{2}$, the corresponding data show that the absolute value of the
carrier density of MgB$_{2}$ is about two orders larger than that of YBa$%
_{2} $Cu$_{3}$O$_{7},$ as suggested by experimental studies. \cite{kang} The
analytical result confirms the existence of the hexagonal vortex lattice
[SH2(2,1,2) phase] in MgB$_{2}$.\cite{eskildsen} Figure \ref{mgb1} shows the
vortex line (charge stripe) structures in the superconducting plane (B
plane) of the SH2(2,1,2) inclined hexagonal vortex lattice. The hexagonal
vortex lattice possessing similar structure as Fig. \ref{4phase}(c) can be
experimentally observed in XZ plane, it should be noted that vortex lines
are non-perpendicular to the XZ plane (with a included angle 60$^{0}$). In
addition, MgB2 is a non-pseudogap superconductor due to a much higher charge
carrier density, as shown in Table \ref{table1}.

\begin{figure}[tbp]
\begin{center}
\resizebox{0.9\columnwidth}{!}{
\includegraphics{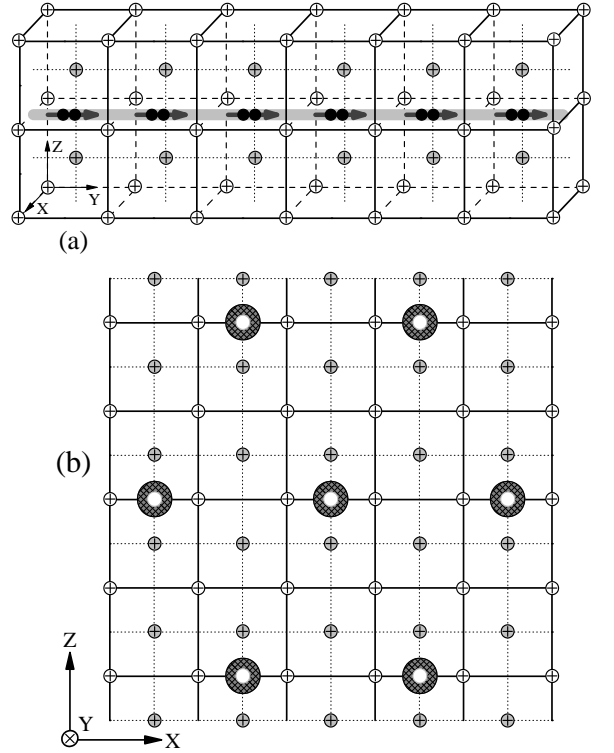}}
\end{center}
\caption{(a) A vortex line (charge stripe) is coherently built up in the
body-centered cubic lattice, (b) a possible hexagonal vortex lattice in the
conventional metallic superconductors. }
\label{metal1}
\end{figure}

To end this section, we would like to present a qualitative interpretation
of the hexagonal vortex lattice and superconducting vortex lines in the pure
metallic superconductors. As shown in Fig. \ref{metal1}, a vortex line
(charge stripe) is coherently built up in the body-centered cubic lattice
because of the real space confinement effect, as shown in Fig. \ref{metal1}
(a). To maintain a stable superconducting phase, different vortex lines
should organize themselves into a periodic vortex lattice, for instance, the
hexagonal vortex lattice of Fig. \ref{metal1} (b).

\subsection{Why the good conductors are non-superconducting}

\begin{figure}[bp]
\begin{center}
\resizebox{1\columnwidth}{!}{
\includegraphics{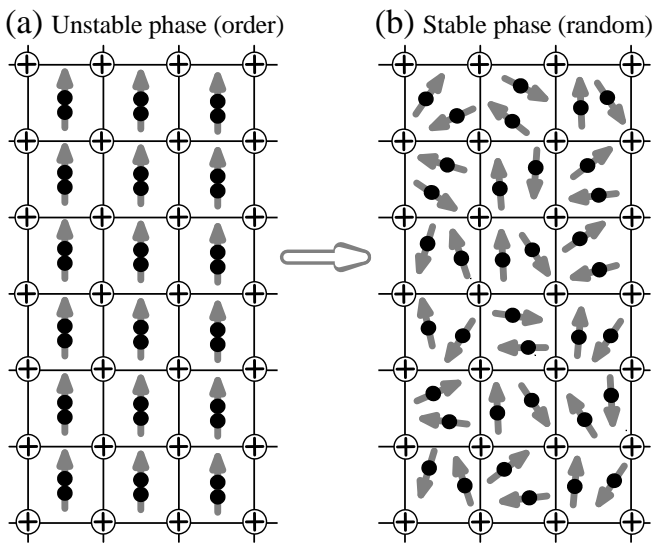}}
\end{center}
\caption{(a) A crowded vortex lattice in a system with high concentrations
of charge carriers, this periodic vortex phase is unstable owing to the
strong electromagnetic interactions between vortex lines. (b) The charge
carriers tend to form in a random phase (non-superconducting) which may be
more stable than the order vortex phase of (a).}
\label{metal}
\end{figure}

The real space confinement pictures (see Figs. \ref{confine1} and \ref%
{4phase}) imply that, to be a superconductor, some periodic and stable
quasi-one-dimensional \textquotedblleft freeways\textquotedblright\ [see
Fig. \ref{metal1} (a)]\ for the superconducting electron pairs should be
built naturally in the system. Any superconducting behavior is always
accompanied by the formation of the vortex lattice in the materials. A
higher superconducting transition temperature only mean the existence of
some more stable \textquotedblleft freeways\textquotedblright\ and vortex
lattice in the superconductor. Therefore, to get higher $T_{c}$
superconductors, the crystal structure and the charge carrier density of the
materials should be taken into account. According to the above discussions,
it is obvious that a appropriate charge carrier density (not too high, not
too low) is helpful for a higher $T_{c}$. Excess charge carrier
concentrations in a material is harmful for superconductivity. As shown in
Fig. \ref{metal}, in a system with high concentrations of charge carriers,
the crowded vortex lattice is unstable owing to the strong electromagnetic
interactions between vortex lines. In this case, the charge carriers are
more likely to be formed in a random and stable phase. This may explain why
the good conductors (for example, Ag, Au, and Cu) and the overdoped high-$%
T_{c}$ superconductors are non-superconducting.

\section{Concluding remarks and further experiments}

Without Hamiltonian, without wave function, without quantum field theory,
our scenario has provided a beautiful and consistent picture for describing
the myriad baffling microphenomena which had previously defied explanation.
The encouraging agreement of our results with the experiments implies a
possibility that our theory would finally open a new window in physics. The
new ideas presented in this paper may change the way we view our world. We
insist that any electronic pairing and superconducting phenomena should
share exactly the same physical reason. We argue that the \textbf{k}-space
quasiparticle picture is very difficult to provide a convincing explanation
of the superconductivity and the famous BCS theory may be incorrect.
Finally, we would like to mention that many results in this paper could be
verified by further experiments.

\section*{Acknowledgments}

The author would like to thank Dr. Kezhou Xie and Dr. Ken C. Lai for many
useful suggestions.

\end{document}